\documentclass{ws-mpla}

\usepackage[super]{cite}
\usepackage{xcolor}
\usepackage[verbose,hypertexnames=false]{hyperref}
\hypersetup{colorlinks=false,allbordercolors=blue,pdfborderstyle={/S/U/W 1}}

\usepackage{graphicx}
\usepackage{tikz}
\usetikzlibrary{shapes, arrows, positioning}

\usepackage{booktabs,array}

\begin{document}

\markboth{Ilham Prasetyo, Bobby Eka Gunara, and Agus Suroso}{Preliminary study on the impact of stress-energy tensor compared to scalar field in Nonminimal Derivative model}

\catchline{}{}{}{}{}

\title{Preliminary study on the impact of stress-energy tensor compared to scalar field in Nonminimal Derivative model
}

\author{Ilham Prasetyo\footnote{Corresponding author. \\
This is a preprint, prior to copyediting, of an article accepted in Modern Physics Letters A on March 31, 2026. The DOI is available at \href{https://doi.org/10.1142/S0217732326420095}{https://doi.org/10.1142/S0217732326420095}}.}

\address{Department of Computer Science, Faculty of Engineering and Technology, Sampoerna University, Pancoran, Jakarta 12780, Indonesia\\
	ilham.prasetyo@sampoernauniversity.ac.id 
}

\author{Bobby Eka Gunara}
\author{Agus Suroso}

\address{Theoretical High Energy Physics Research Division, Institut Teknologi Bandung, Jl. Ganesha 10, Bandung 40132, Indonesia \\
	bobby@itb.ac.id
	surosoagus@itb.ac.id 
}

\maketitle

\begin{history}
\received{(Day Month Year)}
\revised{(Day Month Year)}
\accepted{(Day Month Year)}
\published{(Day Month Year)}
\end{history}

\begin{abstract}
In this article, we report the results of comparing the effect of using trace of stress-energy tensor versus real-valued scalar field in Nonminimal Derivative Coupling gravitation model, respectively denoted as NMDC-T and NMDC-phi. We employ the model into an incompressible star and see the effect of both models NMDC-T and NMDC-phi on the compactness and mass-radius relation. We find that coupling parameters of NMDC-T is less sensitive than NMDC-phi.
\end{abstract}


\ccode{PACS Nos.: 03.65.$-$w, 04.62.+v}

\section{Introduction}
The Nonminimal Derivative Coupling gravitation model is originally a subset of the Fab Four model, a generalization of the Horndeski model.\cite{Charmousis2012} The Horndeski model was established as the most general Lagrangian form of scalar field coupled with curvature tensors whose equation of motions contain only up-to second-order differential term.\cite{Horndeski1974} Initially, the Horndeski model, and subsequently its modified versions including The Fab Four, are proposed as models for explaining the dynamics of the universe.

The Nonminimal Derivative Coupling model (NMDC) has been discussed recently as a candidate for modelling compact stars, for instance see Refs. \refcite{Cisterna2015}, \refcite{Cisterna2016}, and \refcite{Danarianto2025}. The motivation are at least due to two aspects. First aspect is the nondivergence of scalar field constraints that can be obtained using only symmetry arguments.\cite{Hui2013} This leads to a simplification of the equation of motions from the modified Einstein Field Equation (EFE) using constraint equations applied to the scalar’s current density $J^a$. Second aspect is due to the previous investigations of the static black-hole solutions from employing a scalar field $\phi(r)$ that depends only on radial variable\cite{Rinaldi2012} and another one $\phi(t,r)$ that depends on radial and time variables.\cite{Babichev2014} The former leads to a nontrivial black-hole solution that has interesting behavior at some limits. The latter, whose scalar field ansatz is $\phi(t,r)=Qt+F(r)$,  leads to different types of black-holes solutions accompanied with their respective nontrivial form of the scalar field radial term $F(r)$. The latter form is commonly used since one of the black-hole solutions is the Schwarzschild metric.

The NMDC model mentioned above uses real-valued scalar field hence we name it as NMDC-phi. This model, however, suffers from a problematic self-consistency in the scalar field if applied to compact stars and if we set NMDC parameter to be negative\cite{Danarianto2025} $\eta<0$. Inside the star, $F'(r)^2<0$ happens on some range of r inside the compact star for some equation of states, e.g. the neutron star types. This implies complex value of the scalar field, inconsistent to the initial assumption that the scalar field is a real-valued function. Moreover, and somehow related to it, $\eta<0$ is stated to have ghost instability.\cite{Cisterna2015}

This leads us to consider another type of NMDC model where the scalar field is replaced with the trace of the stress-energy tensor $T=g^{ab} T_{ab}$.\cite{Asimakis2023} We will denote this model as NMDC-T. This model was initially used to investigate cosmology with dark energy and checked with a set of cosmological data. It fits with the history of the universe with dark energy epochs. 
The appeal of NMDC-T model compared to NMDC-phi is that it can be used to compact stars without the complex value problem in the scalar field of the NMDC-phi model since NMDC-T with ideal fluid implies $T=-\rho+3P$ which is real-valued, with $\rho$ and $P$ the energy density and pressure, respectively. However, as we shall see below, the modified EFE will contain terms with second covariant derivatives of $T$, implying terms with $P'' (r)$ and $\rho'' (r)$, thus nonsmooth EoS will introduce another problem. We will not discuss this aspect in this article.

In this article, we investigate NMDC-T and NMDC-phi in parallel. In the section 2, we derive the necessary equations of motions (including the modified Tolman-Oppenheimer-Volkoff (TOV) equations) from NMDC-phi model, including boundary conditions and the numerical method. In section 3, we do the same as section 2 but for NMDC-T model. In section 4, we discuss the results from varying the coupling parameters of both NMDC-T and NMDC-phi models and compare them. We restrict our discussion only to incompressible star equation of state (EoS) and we rescale every function and paramaters such that all of them becomes dimensionless in order to keep objective discussion between the two coupling constants.

\section{NMDC-phi model with incompressible EoS}
In this section, we recount the derivation of the equations of motions from NMDC-phi model.\cite{Cisterna2015} The units are all adjusted such that the speed of light is equal to one ($c=1$). We start with action of the following: 
\begin{equation}
	S=\int d^4x \sqrt{-g}  [\kappa(R_a^a-2\Lambda)-(\zeta g_{ab}-\eta G_{ab} ) \nabla^a \phi \nabla^b \phi]+S_m,
\end{equation}
where $\kappa=1/(16\pi G)$ ($G$ is the Newton’s constant), $\Lambda$ the cosmological constant, $\zeta$ the minimal derivative coupling constant, $\eta$ the nonminimal derivative coupling constant, and $S_m$ action terms from matter content. We do not use $R$ as Ricci scalar because we will use it as the variable of the star’s radius. This action, after using variational action, produces equations of motions as follows
\begin{eqnarray}
	\nabla^aT_{ab}&=&0 \text{ with } T_{ab}=-\frac{2}{\sqrt{-g}}\frac{\delta S_m}{\delta g^{ab}}, \\
	\nabla^a J_a&=&0 \text{ with } J_a=(\zeta g_{ab}-\eta G_{ab})\nabla^b \phi, \label{eq:Ja}\\
	G_{ab}&+&\Lambda g_{ab}=H_{ab}+\frac{1}{2\kappa}  T_{ab}, \label{eq:EFE}\\
	H_{ab}&=&\frac{\zeta}{2\kappa} H_{ab}^{(\zeta)}+\frac{\eta}{2\kappa} H_{ab}^{(\eta)}, \\
	H_{ab}^{(\zeta)} &=& \nabla_a \phi \nabla_b \phi - \frac{1}{2} g_{ab} \nabla \phi \nabla \phi, \\
	H^{(\eta)}_{ab} &=& \frac{1}{2} R_c^c \nabla_a \phi \nabla_b \phi - \nabla^c \phi (\nabla_a \phi R_{bc} + \nabla_b \phi R_{ac}) - \nabla^c \phi \nabla^d \phi R_{cabd} \nonumber\\
	&&- \nabla_a \nabla^c \phi \nabla_b \nabla_c \phi + \frac{1}{2} g_{ab} \nabla^c\nabla^d \phi \nabla \phi \nabla_c \nabla_d \phi + \frac{1}{2} g_{ab} (\nabla_c \nabla^c \phi)^2 \nonumber\\
	&&+ \nabla_c \nabla^c \phi \nabla_a \nabla_b \phi + \frac{1}{2} G_{ab} \nabla^c \phi \nabla_c \phi + \frac{1}{2} g_{ab} R_{cd} \nabla^c \phi \nabla^d \phi,
\end{eqnarray}
However, instead of using Eq. (\ref{eq:Ja}), one uses constraint $J_r=0$ taken from Ref.~\refcite{Babichev2014}.

We then input inside it the metric ansatz, ideal fluid, and the scalar field
\begin{eqnarray}
	ds^{2} &=& - e^{2A(r)} dt^{2} + e^{2B(r)} dr^{2} + r^{2} d\theta^{2} + r^{2} \sin^{2}(\theta) d\phi^{2}, \\
	T_{ab} &=& \rho(r) U_{a} U_{b} + P(r) \left( g_{ab} + U_{a} U_{b} \right), \\
	\phi &=& Q t + F(r),
\end{eqnarray}
with $U^a$ the normalized 4-velocity for static fluid ($U_a U^a=-1$). We focus only on the nonminimal term and without cosmological constant $\zeta=0=\Lambda$. 

Using {\it Mathematica} with {\it xAct} package,\cite{xAct} one can work out the long derivation which we outline as follows. Using $J_r=0$, we obtain
\begin{equation}
	A' = \frac{e^{2B} - 1}{2r}, \label{eq:Aeta}
\end{equation}
where we denote derivatives with respect to $r$ as primes. It turns out that $J_r = 0$ gives $H_{tr} = 0$. Inputting Eq. (\ref{eq:Aeta}) into Eq. (\ref{eq:EFE}) then extract the $rr$-component, we have
\begin{equation}
	F'^2 = e^{-2A} \left( r^2 P\, e^{2(A+B)} + 2\eta Q^2 \left( e^{2B} - 1 \right) \right) / (2\eta).
\end{equation}
For convenience, one can obtain $F^{I\,\prime\prime}$ from applying a derivative with respect to $r$ into the $rr$-component of Eq. (\ref{eq:EFE}). From Eq. (\ref{eq:Ja}), we have
\begin{equation}
	P' = -A'(P + \rho).
\end{equation}
Lastly, from inputting $A'$, $F^{I\,\prime 2}$, $F''$, and $P'$ into the $tt$-component of Eq. (\ref{eq:EFE}), we obtain
\begin{eqnarray}
		B' &=& \left\{ 6\, r^2 P\, e^{2(A+B)} + 2 \left( e^{2B} - 1 \right) \left( 3\eta Q^2 - 2\kappa\, e^{2(A+B)} \right) + r^2 \left( e^{2B} + 1 \right) \rho\, e^{2(A+B)} \right\} \nonumber\\
		&&\times \left[ 2r \left( r^2 P\, e^{2(A+B)} + 4\kappa\, e^{2(A+B)} - 6\eta Q^2 \right) \right]^{-1},
		\label{eq:Beta}
\end{eqnarray}

Notice that $A'$ is obtained from conservation of the scalar current, in contrast to the usual TOV equation derived from general relativity (TOV-GR) where $A'$ is from the $tt$-component of the Einstein field equation. This implies\cite{Cisterna2015} the result of this model not approaching the result of TOV-GR in the limit of $\eta \to 0$.

Another aspect related to numerical calculation is discussed in the following. In the literatures, the dimension of the scalar field $\phi$ is usually not discussed. One can set its dimension arbitrarily and rescue the terms in the Lagrangian by introducing the coupling constant with their appropriate dimensions such that the overall term has dimension $\frac{M}{L^3}$. In our case here, we will set the scalar field’s dimension to be $\frac{M}{L^3}$ the same as the trace of stress-energy tensor, following the construction in NMDC-T model. This choice implies the dimension of $\zeta$ and $\eta$ to be $\frac{L^5}{M}$ and $\frac{L^7}{M}$, respectively. For completeness, $\kappa$ has dimension of $\frac{M}{L}$.

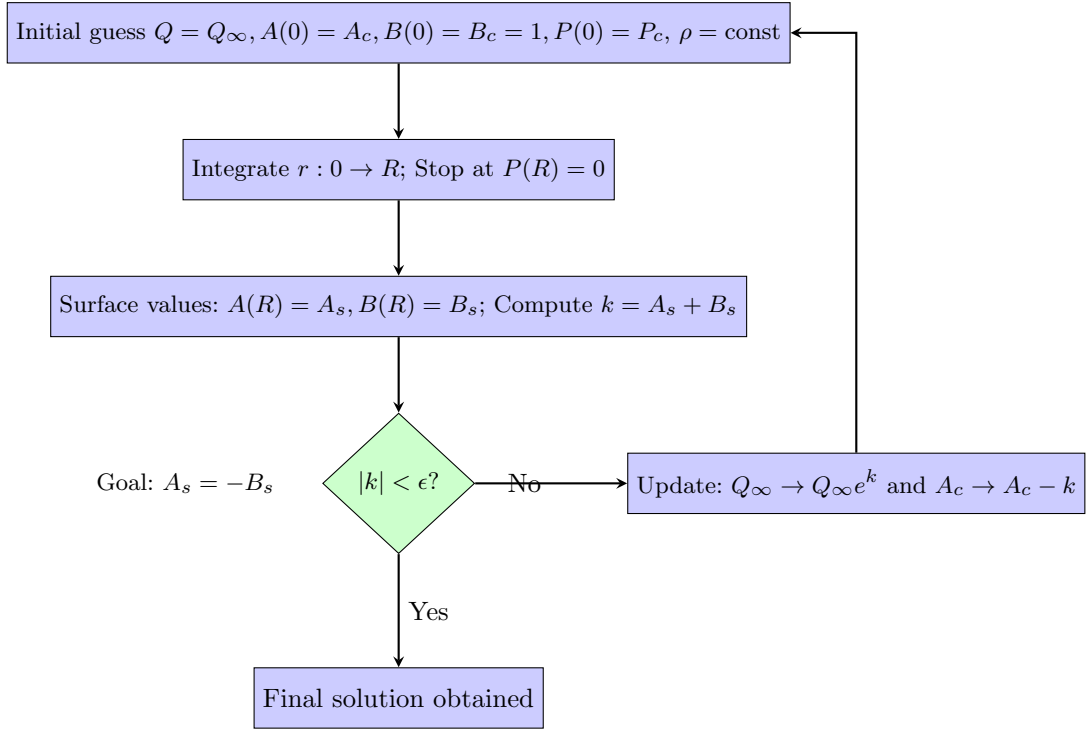
\begin{figure}[t]
\centering
\begin{tikzpicture}[
    node distance=1cm and 0.8cm,
    box/.style={rectangle, minimum width=3cm, minimum height=0.8cm, text centered, draw=black, fill=blue!20},
    decision/.style={diamond, minimum width=2cm, minimum height=1cm, text centered, draw=black, fill=green!20},
    arrow/.style={thick,->,>=stealth}
]

\node (start) [box, font=\small] {Initial guess $Q=Q_\infty, A(0)=A_c, B(0)=B_c=1, P(0)=P_c$,  $\rho=\text{const}$};
\node (integrate) [box, below=of start, font=\small] {Integrate $r: 0 \to R$; Stop at $P(R)=0$};
\node (check) [box, below=of integrate, font=\small] {Surface values: $A(R)=A_s, B(R)=B_s$; Compute $k = A_s + B_s$};
\node (decide) [decision, below=of check,  font=\small] {$|k|<\epsilon$?};
\node (update) [box, right=2cm of decide,  font=\small] {Update:  $Q_\infty \to Q_\infty e^{k}$ and $A_c \to A_c - k$};
\node (final) [box, below=1.5cm of decide] {Final solution obtained};

\draw [arrow] (start) -- (integrate);
\draw [arrow] (integrate) -- (check);
\draw [arrow] (check) -- (decide);
\draw [arrow] (decide) -- node[left] {No} (update);
\draw [arrow] (update) |- (start);
\draw [arrow] (decide) -- node[right] {Yes} (final);

\node [right=0.5cm of start, align=left, font=\small] {};
\node [right=0.5cm of integrate, align=left, font=\small] {};
\node [right=0.5cm of check, align=left, font=\small] {};
\node [left=0.5cm of decide, align=right, font=\small] {Goal: $A_s=-B_s$};

\end{tikzpicture}
\caption{ {Simplified flowchart of the shooting method for NMDC-phi model.}}
\label{fig:shooting_method}
\end{figure}

 {We do numerical calculation from the center of the star ($r\sim0$) to the surface of the star ($r=R$). The flowchart in Figure~\ref{fig:shooting_method} illustrates the process of it. We employ the shooting method in the numerical integration with boundary condition matching to the Schwarzschild exterior. The algorithm iteratively adjusts the value of $A(0)=A_c$ and $Q=Q_\infty$, where the latter is observed by observer at faraway distances, until the surface condition $A(R) = -B(R)$ is satisfied.}


As we shall see by comparison to NMDC-T model, Eqs. (\ref{eq:Aeta})-(\ref{eq:Beta}) does not employ any expansion with respect to small values of $\eta$. However, it turns out that by analysis around the center of the star, $\eta$ is constrained by
\begin{eqnarray}
	\eta<{2\kappa\over 3}  {e^{2A_c}\over Q^2} \text{ if }\eta>0\text{, and } \label{eq:etapos}
	\\
	|\eta|<{6 P_c \kappa\over (2\rho_c-3P_c ) }  {e^{2A_c}\over Q^2}, \text{ if }\eta<0, \label{eq:etaneg}
\end{eqnarray}
where $\rho(r\to 0)=\rho_c$  {and $P(r\to 0)=P_c$}.

\section{NMDC-T model with incompressible EoS}

In this section, we will derive the needed equations of motions from NMDC-T model. The action has the following form\cite{Asimakis2023}
\begin{eqnarray}
	S=\int d^4 x \sqrt{-g}  [\kappa(R_a^a-2\Lambda)+\alpha G_{ab}  T^{ab}+\beta G_{ab}  \nabla^a T^c_c \nabla^b T^d_d]+S_m  ,
\end{eqnarray} 
with $S_m=\int d^4 x \sqrt{-g} L_m$ and $T_{ab}=-{2\over \sqrt{-g}} {\delta S_m\over \delta g^{ab}}=g_{ab} L_m-2 {\delta L_m\over \delta g^{ab}}$. Using variational action and $T\equiv T_c^c$, one obtains
\begin{eqnarray}
	G_{a b}&+&\Lambda g_{a b}=\frac{1}{2 \kappa}\left(T_{a b}+\alpha T_{a b}^{(\alpha)}+\beta T_{a b}^{(\beta)}\right) \\
	T_{a b}^{(\alpha)}&=&-R_{c}^{c} T_{a b}+G^{c d}\left(g_{a b} T_{c d}-2 g_{b c} T_{d a}-2 g_{a c} T_{d b}\right)+R_{a b} T_{c}^{c}-2 \Xi_{a b} \nonumber\\
	&& \quad -\nabla_{b} \nabla_{a} T+\nabla_{d} \nabla_{a} T_{b}^{d}+\nabla_{d} \nabla_{b} T_{a}^{d}-g_{a b} \nabla_{d} \nabla_{c} T^{c d}+g_{a b} \nabla_{d} \nabla^{d} T \nonumber\\
	&& \quad -\nabla_{c} \nabla^{c} T_{a b},
\end{eqnarray}
\begin{eqnarray}
    T_{a b}^{(\beta)}&=&-R_{e}^{e} \nabla_{a} T \nabla_{b} T-2 G_{b}^{c} \nabla_{a} T \nabla_{c} T-2 G_{a}^{c} \nabla_{b} T \nabla_{c} T+G^{c d} g_{a b} \nabla_{c} T \nabla_{d} T \nonumber\\
	&& \quad +g_{a b} R^{c d} \nabla_{c} T \nabla_{d} T+4 G^{c d} T_{a b} \nabla_{d} \nabla_{c} T+4 G^{c d} \Theta_{a b} \nabla_{d} \nabla_{c} T \nonumber\\
	&& \quad +R_{a b} \nabla_{d} T \nabla^{d} T-2\left(\nabla_{d} \nabla_{b} T\right)\left(\nabla^{d} \nabla_{a} T\right)+2\left(\nabla_{b} \nabla_{a} T\right)\left(\nabla_{e} \nabla^{e} T\right) \nonumber\\
	&& \quad +g_{a b}\left(\nabla_{e} \nabla_{d} T\right)\left(\nabla^{e} \nabla^{d} T\right)-g_{a b}\left(\nabla_{d} \nabla^{d} T\right)\left(\nabla_{f} \nabla^{f} T\right) \nonumber\\
	&& \quad -4 R_{a d b f} \nabla^{d} T \nabla^{f} T-R^{i}{ }_{b a d} \nabla^{d} T \nabla_{i} T-R^{i}{ }_{a b d} \nabla^{d} T \nabla_{i} T, \\
	\Theta_{a b}&=&g^{c d} \frac{\delta T_{c d}}{\delta g^{a b}}=-2 g^{c d} \frac{\delta^{2} L_{m}}{\delta g^{a b} \delta g^{c d}}+g_{a b} L_{m}-2 T_{a b}, \label{eq:Theta}\\
	\Xi_{a b}&=&G^{c d} \frac{\delta T_{c d}}{\delta g^{a b}}=-2 G^{c d} \frac{\delta^{2} L_{m}}{\delta g^{a b} \delta g^{c d}}-G_{a b} L_{m}+\frac{1}{2} G^{c d} g_{c d}\left(g_{a b} L_{m}-T_{a b}\right). \label{eq:Xi}
\end{eqnarray}
Lastly, from the Bianchi identity, we have a constraint
\begin{equation}
	\nabla^{a}\left(T_{a b}+\alpha T_{a b}^{(\alpha)}+\beta T_{a b}^{(\beta)}\right)=0
\end{equation}

One technical aspect must be noted here. Notice that we need another functional derivative $\frac{\delta T_{c d}}{\delta g^{a b}}$ to obtain Eqs. (\ref{eq:Theta})-(\ref{eq:Xi}). Here we use $T_{c d}=\rho U_{c} U_{d}+P\left(g_{c d}+U_{c} U_{d}\right)$ where $U^{a}$ is the normalized fluid's 4-vector velocity $U_{a} U^{a}=-1$. In explicit case we consider here, the only nonzero term is $U^{t}=e^{-A}$. We assume that $U^{a}$ is the original one rather than $U_{a}$, hence $U_{a}=U^{b} g_{a b}$. Since both energy density and pressure are dependent only on $r$, then $\delta g^{a b}$ will only come from $g_{c d}$ and $U_{c} U_{d}$. Moreover, one can choose either $L_{m}=-\rho$ or $L_{m}=P$ since both can lead to the same $T_{c d}$.\cite{Brown1993} However, for simplicity, we use $L_{m}=P$ following Ref. 9. This then give us
\begin{eqnarray}
	\Theta_{c d}&=&-g_{c d} P-2 U_{c} U_{d}(P+\rho) \\
	\Xi_{c d}&=&-G_{c d} P+\frac{1}{2} G_{a b} g^{a b} U_{c} U_{d}(P+\rho)
\end{eqnarray}
Notice that there are second derivatives of $T$ inside $T_{a b}^{(\alpha)}$ and $T_{a b}^{(\beta)}$. Since we use ideal fluid, $T=-\rho+3 P$, so we can expect terms with $\rho^{\prime \prime}$ and $P^{\prime \prime}$. This leads us to a problem, i.e., if we find the expression of $P^{\prime \prime}$ then the form will be $P^{\prime \prime}=\frac{(\cdots)}{\alpha}+\frac{(\cdots)}{\beta}$. In general, this cannot be easily made to approach the usual TOV equation in the limit of either $\alpha \rightarrow 0$ or $\beta \rightarrow 0$.

Therefore, we attempt on doing a recursion method to ensure that the resulting equations will approach the usual TOV equations when $\alpha \rightarrow 0$ or $\beta \rightarrow 0$. The recursion method is outlined in the following steps. First, we follow the derivation of the usual TOV equation starting by arranging the resulting components of the modified Einstein field equation to be, e.g.
\begin{eqnarray}
P'=(\text{GR terms})+\alpha P_1 (P,P',P'',\rho,\rho',\rho'',A,A',A'',B,B',B'',r)\nonumber\\+\beta P_2 (P,P',P'',\rho,\rho',\rho'',A,A',A'',B,B',B'',r),
\end{eqnarray}
where $P_1$ and $P_2$ are the correction terms from nonminimal coupling and nonminimal derivative coupling terms, respectively. $B'$ and $A'$ also have a similar form. The higher order derivative terms contain $P'', \rho'', A'',$ and $B''$. This does not apply any approximation yet. 

Second, we set our target to obtain the equations that only contain correction up to first order of $\alpha$ and $\beta$. By this choice, we do derivative of GR terms with respect to r, i.e., $P'$ to obtain $P''$, and so on. 

Third, these will be inputted into the modified terms to obtain, e.g.,
\begin{eqnarray}
	P'=(\text{GR terms})+\alpha P_{1,0} (P,\rho,\rho',\rho'',A,B,r)\nonumber\\
	+\beta P_{2,0} (P,\rho,\rho',\rho'',A,B,r)+O(\alpha^2,\beta^2,\alpha\beta),
\end{eqnarray}
where now the correction terms $P_{1,0}$ and $P_{2,0}$ only contain $P,\rho,\rho',\rho'',A,$ and $B$ and the last term with second and higher orders are ignored. We cannot eliminate $\rho'$ and $\rho''$ because they depend on the inputted equation of state.
 
Since we intend to only focus on the nonminimal terms, we set $\Lambda=0=\alpha$. This results in the following list of equations\footnote{with $\rho'\equiv d\rho/dP$ and $\rho''\equiv d^2\rho/dP^2$}:
\begin{align}
B'(r)
&= -\frac{e^{2B}-1}{2r}
   + \frac{\rho r e^{2B}}{4\kappa}
   - \frac{\alpha}{32\kappa^3 r}
\biggl\{
        r^4 e^{4B} P^3 (\rho' + 2)
\nonumber\\
&\quad\quad
      + r^2 e^{2B} P^2
        \bigl[
           r^2 e^{2B} \rho (\rho' + 2)
           + 4\kappa (e^{2B}-1)\rho'
           + 8\kappa (e^{2B}-4)
        \bigr]
\nonumber\\
&\quad\quad
      + 4\kappa \rho
        \bigl[
            \kappa (e^{2B}-1)^2 (\rho' + 2)
            - r^2 e^{2B}\rho
        \bigr]
\nonumber\\
&\quad\quad
      + 4\kappa P
        \bigl\{
            r^2 e^{2B} \rho
            \bigl[
                (e^{2B}-1)\rho' + 2e^{2B}-11
            \bigr]
            + \kappa (e^{2B}-1)^2 (\rho' + 2)
        \bigr\}
\biggr\} \nonumber\\
&\quad
    -\frac{\beta e^{-2B}(P+\rho)^2}{128\kappa^4 r^3}
\Biggl\{
   2 r^6 e^{6B} P^4 (\rho' - 5)\rho''  + 4\kappa^2 (e^{2B}-1)(\rho'-3)
\nonumber \\
&\qquad
  \times \Biggl[
      \rho
      \biggl(
         r^2 e^{2B}(e^{2B}-5)\rho'
         + 4\kappa (e^{2B}-1)^2 \rho''
         + r^2 e^{2B}(19 - 7e^{2B})
      \biggr)
\nonumber\\
&\quad\quad\quad
      + 4\kappa (e^{2B}-1)(\rho'-3)
      \biggl[
         (e^{2B}-1)\rho' + e^{2B} + 2
      \biggr]
   \Biggr]
\nonumber\\
&\quad
   + 4\kappa P
   \Biggl[
      2\kappa (e^{2B}-1)^2(\rho'-3)
      \biggl(
         3 r^2 e^{2B}\rho'^2
         - 7 r^2 e^{2B}\rho'
         + 2\kappa(e^{2B}-1)\rho''
         - 16 r^2 e^{2B}
      \biggr)
\nonumber\\
&\quad\quad
      + r^2 e^{2B}\rho
      \Biggl\{
         \rho'
         \biggl[
             6\kappa (e^{2B}-1)^2 \rho''
             - 2 r^2 e^{2B}(5e^{2B}-13)
         \biggr]
\nonumber\\
&\quad\quad\quad
         + r^2 e^{2B}(e^{2B}-3)\rho'^2
         - 22\kappa (e^{2B}-1)^2 \rho''
         + 3 r^2 e^{2B}(7 e^{2B}-17)
      \Biggr\}
   \Biggr]
\nonumber\\
&\quad
   + 2 r^4 e^{4B} P^3
   \Biggl[
      \rho''
      \biggl(
         -5 r^2 e^{2B}\rho
         - 26\kappa (e^{2B}-1)
      \biggr)   \nonumber \\
&\quad\quad
      + \rho'
      \Biggl\{
          \rho''
          \bigl(
              r^2 e^{2B}\rho
              + 6\kappa (e^{2B}-1)
          \bigr)
          - r^2 e^{2B}
      \Biggr\}
      + r^2 e^{2B}\rho'^3
      - 6 r^2 e^{2B}\rho'^2
      + 30 r^2 e^{2B}
   \Biggr]
\nonumber\\
&\quad
   + P^2
   \Biggl\{
      4\kappa r^2 e^{2B}
      \Biggl[
         \rho'
         \Bigl\{
             6\kappa (e^{2B}-1)^2 \rho''
             + r^2 e^{2B}(e^{2B}+29)
         \Bigr\}
\nonumber\\
&\quad\quad
         + 3 r^2 e^{2B}(e^{2B}-1)\rho'^3
         - r^2 e^{2B}(17 e^{2B}-14)\rho'^2
         - 22\kappa (e^{2B}-1)^2\rho''
         + 3 r^2 e^{2B}(23 e^{2B}-44)
      \Biggr]
\nonumber\\
&\quad\quad
      + r^4 e^{4B}\rho
      \Biggl[
         -2\rho'
         \biggl(
             5 r^2 e^{2B}
             - 6\kappa (e^{2B}-1)\rho''
         \biggr)  + r^2 e^{2B}\rho'^2
         - 52\kappa (e^{2B}-1)\rho''
         + 21 r^2 e^{2B}
      \Biggr]
   \Biggr\}
\Biggr\}      
\label{eq:Bbeta}
\end{align}
\begin{align}
A'(r)
&= \frac{r e^{2B} P}{4\kappa}
   + \frac{e^{2B}-1}{2r}
\notag\\
&\quad
   + \frac{\alpha}{8\kappa^2 r}
   \Biggl\{
      -\frac{1}{4\kappa}(\rho+P)
      \Biggl[
         r^2 e^{2B} P
         + 2\kappa (e^{2B}-1)
      \Biggr]
\notag\\
&\qquad\qquad\qquad\times
      \Biggl[
         r^2 e^{2B} P
         + 2\kappa (e^{2B} - 2\rho' + 1)
      \Biggr]
      - r^2 e^{2B} P\,\rho
      + r^2 e^{2B} P^2
   \Biggr\}
\notag\\
&\quad
   - \frac{\beta (4\kappa + 3 r^2 P)}{128\kappa^4 r^3}
     (\rho + P)^2
     \Biggl[
        r^2 e^{2B} P + 2\kappa (e^{2B}-1)
     \Biggr]^2
     (\rho' - 3)^2,
\\[1em]
P'(r)
&= -\frac{(\rho+P)}{4\kappa r}
     \Biggl[
        r^2 e^{2B} P
        + 2\kappa (e^{2B}-1)
     \Biggr]
\notag\\
&\quad
   + \frac{\alpha}{8\kappa^2 r}
   \Biggl\{
      \rho (\rho + P)
      \Biggl[
          r^2 e^{2B} P
          + 2\kappa (e^{2B}-1)
      \Biggr]
      (\rho' - 1)
   \Biggr\}
\notag\\
&\quad
   - \frac{\beta e^{-2B}(\rho+P)^2}{64\kappa^4 r^3}
     \Biggl[
         r^2 e^{2B} P + 2\kappa (e^{2B}-1)
     \Biggr]
     (\rho' - 1)
\notag\\
&\qquad\times
   \Biggl\{
      r^4 e^{4B} P^4 \rho''
\notag\\
&\qquad\quad
      + r^2 e^{2B} P^3
        \Biggl(
           \rho''
           \Bigl[
             r^2 e^{2B}\rho
             + 4\kappa (e^{2B}-1)
           \Bigr]
           + r^2 e^{2B} (\rho'^2 - \rho' - 6)
        \Biggr)
\notag\\
&\qquad\quad
      + P^2
      \Biggl[
         r^2 e^{2B}\rho
         \Biggl(
            r^2 e^{2B}\rho'
            + 4\kappa (e^{2B}-1)\rho''
            - 3 r^2 e^{2B}
         \Biggr)
\notag\\
&\qquad\qquad
         + 4\kappa
         \Biggl(
            r^2 e^{2B}(e^{2B}-1)\rho'^2
            - r^2 e^{2B}(e^{2B}+2)\rho'
            + \kappa (e^{2B}-1)^2\rho''
            - 3 r^2 e^{2B}(2 e^{2B}-5)
         \Biggr)
      \Biggr]
\notag\\
&\qquad\quad
      + 4\kappa P
      \Biggl[
         \rho
         \Bigl(
            r^2 e^{2B}(e^{2B}-2)\rho'
            + \kappa (e^{2B}-1)^2\rho''
            - 3 r^2 e^{2B}(e^{2B}-2)
         \Bigr)
\notag\\
&\qquad\qquad
         + \kappa (e^{2B}-1)^2(\rho'^2 - \rho' - 6)
      \Biggr]
\notag\\
&\qquad\quad
      + 4\kappa^2 (e^{2B}-1)^2 \rho (\rho' - 3)
   \Biggr\}.
\label{eq:Pbeta}
\end{align}

For completeness, we mention the dimension of the coupling parameters in NMDC-T model and the numerical calculation scheme. Since $T_{ab}$ and $T$ have the same dimension $M/L^3$, then $\alpha$ and $\beta$ have dimension $L^2$ and $L^7/M$, respectively. This guarantees the same dimension for $\eta$ in NMDC-phi and for $\beta$ in NMDC-T. Moreover, to satisfy boundary conditions at the surface of the star, we can use the same algorihtm as in the previous section. However, here it is much simpler. Since we have no time-dependence, the numerical calculation needs repeating after only shifting shift the value $A_c$ into another value $A_c\to A_c-k$.  {This is similar to Fig.~\ref{fig:shooting_method} but without the need to use $Q_\infty$}

Notice that since the calculation only investigate up to 1st order correction, we may obtain very large values of $\beta$, unlike $\eta$ which has constraints (see Eqs. (\ref{eq:etapos})-(\ref{eq:etaneg})).

\section{Results and discussion}

The units used for our numerical calculations is called the natural units, where $\hbar c=1=197.33$ MeV.fm.  {This convention needs some conversion if we need the result shown in SI units. We summarize these in Table~\ref{tab:nmdc_conversions}. The quantities not inside square bracket are parameters with a fixed value.} Since we set the units of the scalar field be the same as the unit of trace of the stress-energy tensor $[\phi]=[T]=[P]=$MeV/(fm$^3$), this implies that $[\eta/\kappa]=[\beta/\kappa]=[R]/[G^{ab} \nabla_a T \nabla_b T] =[1/P'^2 ]=$m$^2$/(MeV fm$^{-3}$ )$^2$. In the following, we show both $\eta/\kappa$ and $\beta/\kappa$ values without units for brevity  {noting that both had units of m$^2$/(MeV fm$^{-3}$ )$^2$}. 

\begin{table}[t]
\centering
\caption{ {Units and conversion factors from the natural units to SI units for all variables.}}
\label{tab:nmdc_conversions}
\vspace{0.1cm}
\begin{tabular}{l l l}
\toprule 
\textbf{Quantity} & \textbf{Natural Units} & \textbf{SI Units} \\
{} & $\hbar c=1 = 197.33$ MeV$\cdot$fm & -- \\
\midrule
$G$ & $1.325\times 10^{-12}$ fm/MeV (fm$^2$)/m$^2$ & $6.6726\times 10^{-11}$ m$^3$/(kg$\cdot$s$^2$) \\
$M_\odot$ & $1.1155\times 10^{15}$ MeV m$^3$/fm$^3$ & $1.98892\times 10^{30}$ kg \\
$\kappa=c^4/(16\pi G)$ & $1.50146 \times 10^{10}$ (MeV/fm) m$^2$/fm$^2$ & $2.40773\times 10^{42}$ kg.m/s$^2$ \\  
$[P]$ & MeV/fm$^3$ & $1.6022\times 10^{33}$ dyne/cm$^2$ \\
$[\rho]$ & MeV/fm$^3$ & $1.7827\times 10^{12}$ g/cm$^3$ \\
$[r]$ & m & m \\
$[\phi]$, $[T]$ & assumed to be the same as $[P]$ & {}\\
$[\eta/\kappa]$, $[\beta/\kappa]$ & m$^2$/(MeV$\cdot$fm$^{-3}$)$^2$ & $3.896\times 10^{-63}$ cm$^4$/dyne \\
\bottomrule
\end{tabular}
\end{table}

Here we show the results when we set $\rho=1000$ MeV/fm$^3$. We do variations of both $\eta/\kappa$ and $\beta/\kappa$ and see the result on the compactness $GM/R$. We also show the result of both MR relations and central pressure $P_c$ vs compactness. 

\begin{figure}
	\centering
	\includegraphics[width=0.6\linewidth]{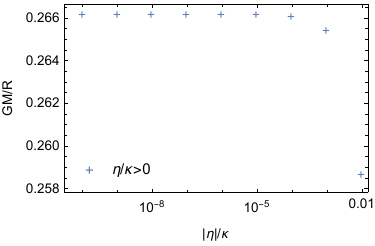}
	\caption{NMDC-phi variation of parameter $\eta>0$.}
	\label{fig:plot0nmdcphi}
\end{figure}

\begin{figure}
	\centering
	\includegraphics[width=0.6\linewidth]{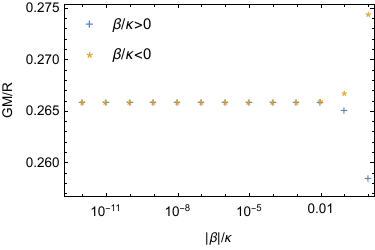}
	\caption{ {NMDC-T} variation of parameter $\beta>0$  {and $\beta<0$}.}
	\label{fig:plot0nmdct}
\end{figure}

As we can see in Figs. \ref{fig:plot0nmdcphi} and \ref{fig:plot0nmdct}, the compactness decreases as the NMDC parameter is increased into positive direction. We do not show results from $\eta<0$ from NMDC-phi because there is a pathology on the value of the scalar field $\phi=Qt+F(r)$, i.e., its value becomes complex because $F'(r)<0$ at some region of $r$ although its MR curve is shifted upward, indicating higher mass. This higher mass is similar to the behaviour of NMDC-T. However, this pathology does not exists for NMDC-T because $T=3P-\rho$ are guaranteed to be real valued, therefore an increase in mass and compactness can be achieved by setting $\beta<0$.

\begin{figure}
	\centering
	\includegraphics[width=1.\linewidth]{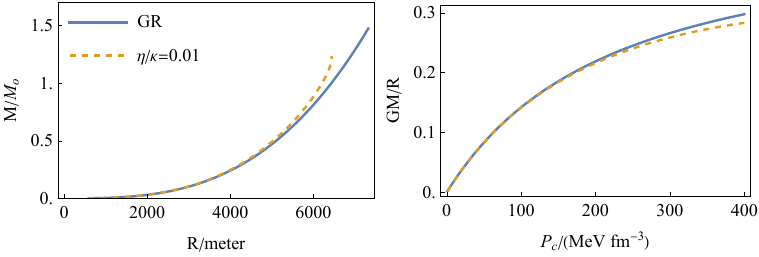}
	\caption{MR relation  {(left)} and central  {pressure} vs. compactness relation  {(right)} for NMDC-phi.}
	\label{fig:plotnmdcphi}
\end{figure}
\begin{figure}
	\centering
	\includegraphics[width=1.\linewidth]{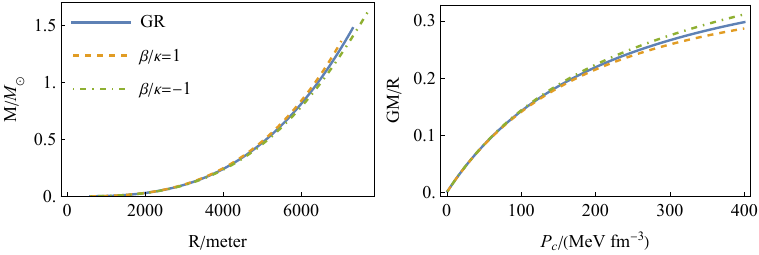}
	\caption{MR relation  {(left)} and central  {pressure} vs. compactness relation  {(right)} for NMDC-T.}
	\label{fig:plotnmdct}
\end{figure}

However, as shown in Figs. \ref{fig:plotnmdcphi} and \ref{fig:plotnmdct}, the shift on the central pressure vs compactness curves is comparable to each other, while NMDC-phi gives a more noticeable shift on MR curves than NMDC-T. Moreover, NMDC-T parameter needs much larger values compared to NMDC-phi, i.e., $\eta/\kappa=0.01$ as opposed to $\beta/\kappa=1$. This perhaps indicates that the results are due to the linear expansion on NMDC-T equations (see Eqs. (\ref{eq:Bbeta})-(\ref{eq:Pbeta}) compared to Eqs. (\ref{eq:Aeta})-(\ref{eq:Beta})). Therefore, higher-order terms with respect to $\alpha$ and $\beta$ perhaps will lead to more noticeable shift.

Notice also that due to constant energy density, the contribution from $\rho'(P)$ and $\rho''(P)$ are not present. These terms may lead to more interesting results since many realistic EoSs have non-continouous $\rho(P)$ especially on some regions of small pressures.

\section{Conclusion}
Here we report the results of comparing the effect of using trace of stress-energy tensor $T=T_a^a$ versus real-valued scalar field $\phi$ in Nonminimal Derivative Coupling gravitation model, respectively denoted as NMDC-T and NMDC-phi. We employ the model into an incompressible star and see the effect of both models NMDC-T and NMDC-phi on the compactness and mass-radius relation. The results are described as follows. First, the NMDC-T model cannot yet be made into numerically solvable differential equations unlike NMDC-phi equations that can be obtained without any expansion. Second, when we increase both NMDC-phi and NMDC-T parameters into more positive values, their behavior are the same, that is, both decrease the incompressible star’s mass. However, NMDC-T parameters can be negative valued, resulting in higher mass than GR results without any pathology unlike NMDC-phi. Third, the result from NMDC-T needs much higher values of its parameter (around 100 times) compared to NMDC-phi, indicating the limit of linear expansion on NMDC-T differential equations. Lastly, the choice of using incompressible star makes derivatives of energy density vanish, therefore using other forms of EoS may lead to more interesting deviations from GR.


 {Now we ask: can we favor NMDC-T over NMDC-phi from a physical or experimental view? From our results, we note two important points. (a) NMDC-phi does not need expansion with respect to small $\eta/\kappa$ unlike NMDC-T, where we only keep up to $O(\beta/\kappa)$. (b) Both NMDC-phi with $\eta<0$ and NMDC-T with $\beta<0$ can increase masses while NMDC-phi with $\eta>0$ and NMDC-T with $\beta>0$ decrease the masses. From point (b), however, Ref.~\refcite{Danarianto2025} shows that NMDC-phi with $\eta< 0$ produce solutions where the (real) scalar field are sometimes complex valued inside the star, especially notable for highly dense stars. This is the reason why we do not show results from NMDC-phi $\eta<0$. This makes NMDC-phi an unlikely solution to explain the Tolman–Oppenheimer–Volkoff limit 2.25 $M_\odot$ from the GW170817 event~\cite{Fan:2023spm,Rezzolla:2017aly,LIGOScientific:2017vwq}. For context, a typical neutron star has mass around 1.4 $M_\odot$ and radius on the order of 10 km. On the contrary, NMDC-T may still do this because $\beta < 0$ solutions contain no complications because both $\rho$ and $P$ are all real valued. However, from point (a), complications may arise if we include higher order terms $O[(\beta/\kappa)^2]$. We shall address this nonlinearity aspect of NMDC-T in a future work. Hence, the short answer to the question in the beginning of this paragraph is yes, because NMDC-T with $\beta<0$ and with correction term ignoring $O[(\beta/\kappa)^2]$ may answer why the Tolman–Oppenheimer–Volkoff limit can be so high.}



\section*{Acknowledgments}
IP is funded by Internal Research Grant from Center of Research and Community Service, Sampoerna University No. 010/IRG/SU/AY.2025-2026. BEG and AS acknowledge Hibah Riset ITB 2025 No. 841/IT1.B07.1/TA.00/2025.

\section*{ORCID}
\noindent Ilham Prasetyo - \url{https://orcid.org/0000-0002-0879-7777}

\noindent Bobby Eka Gunara - \url{https://orcid.org/0000-0002-6529-7664}

\noindent Agus Suroso - \url{https://orcid.org/0000-0002-9367-7613}



\begin{thebibliography}{0}

\bibitem{Charmousis2012}
C. Charmousis, E. J. Copeland, A. Padilla, and P. M. Saffin, \textit{Phys. Rev. D} \textbf{85}, 104040 (2012).

\bibitem{Horndeski1974}
G. W. Horndeski, \textit{Int. J. Theor. Phys.} \textbf{10}, 363-384 (1974).

\bibitem{Cisterna2015}
A. Cisterna, T. Delsate, and M. Rinaldi, \textit{Phys. Rev. D} \textbf{92}, no.4, 044050 (2015).

\bibitem{Cisterna2016}
A. Cisterna, T. Delsate, L. Ducobu, and M. Rinaldi, \textit{Phys. Rev. D} \textbf{93}, no.8, 084046 (2016).

\bibitem{Danarianto2025}
M. D. Danarianto, I. Prasetyo, A. Suroso, B. E. Gunara, and A. Sulaksono, \textit{Phys. Dark Univ.} \textbf{48}, 101919 (2025).

\bibitem{Hui2013}
L. Hui, and A. Nicolis, \textit{Phys. Rev. Lett.} \textbf{110}, 241104 (2013).

\bibitem{Rinaldi2012}
M. Rinaldi, \textit{Phys. Rev. D} \textbf{86}, 084048 (2012).

\bibitem{Babichev2014}
E. Babichev and C. Charmousis, \textit{JHEP} \textbf{08}, 106 (2014).

\bibitem{Asimakis2023}
P. Asimakis, S. Basilakos, A. Lymperis, M. Petronikolou, and E. N. Saridakis, \textit{Phys. Rev. D} \textbf{107}, 104006 (2023).

\bibitem{xAct}
J. M. Martin-Garcia et al. \textit{xAct: Efficient tensor computer algebra for the Wolfram Language.} \url{http://www.xact.es}.  Accessed: 2025-11-13.

\bibitem{Brown1993}
J. D. Brown, \textit{Class. Quant. Grav.} \textbf{10}, 1579-1606 (1993).

\bibitem{Fan:2023spm}
 {Y.~Z.~Fan, M.~Z.~Han, J.~L.~Jiang, D.~S.~Shao and S.~P.~Tang, \textit{Phys. Rev. D} \textbf{109}, no.4, 043052 (2024)}

\bibitem{Rezzolla:2017aly}
 {L.~Rezzolla, E.~R.~Most and L.~R.~Weih, \textit{Astrophys. J. Lett.} \textbf{852}, no.2, L25 (2018).}

\bibitem{LIGOScientific:2017vwq}
 {B.~P.~Abbott \textit{et al.} [LIGO Scientific and Virgo], \textit{Phys. Rev. Lett.} \textbf{119}, no.16, 161101 (2017)}



%
%
%



\end{thebibliography}
\end{document}